\documentclass[amssymb,twocolumn,showpacs,prl,aps]{revtex4}
\usepackage{graphicx}
\begin{document}

\title{
        High frequency dynamics in a monatomic glass.
       }

\author{
         T.~Scopigno$^{1}$,
         R.~Di~Leonardo$^{1}$,
         G.~Ruocco$^{1}$
     }
\affiliation{
     $^{1}$Dipartimento di Fisica and INFM, Universit\'a di Roma ``La Sapienza'',
I-00185, Roma, Italy.}
\author{
         A.Q.R.~Baron$^{2}$,
         S.~Tsutsui$^{2}$
         }
\affiliation{
     $^{2}$SPring-8/JASRI, 1-1-1 Kouto, Mikazuki-cho, Sayo-gun, Hyogo-ken 679-5198
     Japan.
     }
\author{
         F. Bossard$^{3}$
         S.N.Yannopoulos$^{3}$
         }
\affiliation{
     $^{3}$
     Foundation for Research and Technology - Hellas, Institute
     of Chemical Engineering and High-Temperature Chemical
     Processes,FORTH-ICE/HT, P.O. Box 1414, GR-265 00 Patras, Greece.
     }

\date{\today}

\begin{abstract}

The high frequency dynamics of glassy Selenium has been studied by
Inelastic X-ray Scattering at beamline BL35XU (SPring-8). The high
quality of the data allows one to pinpoint the existence of a
dispersing acoustic mode for wavevectors ($Q$) of $1.5<Q<12.5$
nm$^{-1}$, helping to clarify a previous contradiction between
experimental and numerical results. The sound velocity shows a
positive dispersion, exceeding the hydrodynamic value by $\approx$
10\% at $Q<3.5$ nm$^{-1}$. The $Q^2$ dependence of the sound
attenuation $\Gamma(Q)$, reported for other glasses, is found to
be the low-$Q$ limit of a more general $\Gamma(Q) \propto
\Omega(Q)^2$ law which applies also to the higher $Q$ region,
where $\Omega(Q)\propto Q$ no longer holds.

\end{abstract}


\maketitle

%]

The possibility of high frequency sound propagation in disordered
systems has stimulated a large number of investigations in the
last twenty years \cite{creta}. In particular, in glassy systems,
the existence of short wavelength acoustic modes can be traced
back to the pioneering numerical simulation works of Grest, Nagel
and Rahman \cite{grest_ljglass}. The details of these high
frequency excitations are closely interwoven with some well known
"anomalies" such as the excess of vibrational density of state
(the so-called Boson Peak) or the thermal conductivity plateau
observed in the $1<T<10$ K region, which mark non-trivial
distinctions from the crystalline state \cite{pohl_review}. In
this respect, glassy selenium constitutes a system of paramount
importance. Not only does it exhibit all the above mentioned
anomalies \cite{pohl_review,buc_se}, but it is also interesting
for its technological impact due to the well known high
sensitiveness to photo-induced effects \cite{KOBOLOV}.

Crystalline selenium exists at least in three different atomic
arrangements. The more stable is the hexagonal form in which the
atoms are distributed along parallel chains with the threefold
spiral along the c axis. The monoclinic phases (a and b) are
metastable and are both characterized by by eight-membered
Se$_8$-rings. Rhombohedral and cubic Se have also been reported.
As the energy difference between the two main crystal phases -
hexagonal and monoclinic - is very small compared to $k_BT_g$
($T_g$=303 K being the glass transition temperature), the
competition between the two local structures endows to selenium an
excellent glass-forming ability; an extremely rare feature for a
monatomic system. For the above reasons, glassy selenium (g-Se)
has been widely investigated in the past by a large variety of
experimental techniques, and also several numerical investigations
have been reported \cite{and_se,phi_se,gar_se,cap_se,for_se}. This
notwithstanding, some crucial aspects related to the features of
the microscopic collective dynamics at wavelength comprising few
atomic units are still unknown. More specifically, numerical
investigations report in this momentum transfer $Q$ region
evidence of collective modes up to frequencies as high as $8-10$
meV, characterized by a well defined phonon-like dispersion curve
\cite{gar_se} . However, based on recent inelastic neutron
scattering (INS) experiments, a localization threshold of $3-4$
meV ($Q \approx 3$ nm$^{-1}$) was proposed, above which the
longitudinal acoustic dynamics became strongly overdamped
\cite{for_se}.

In this letter we present the first direct determination of the
coherent dynamic structure factor by means of Inelastic X-ray
Scattering (IXS) in g-Se. The extremely favorable
inelastic/elastic intensity ratio allows for a reliable analysis
of the collective dynamics. We find evidence for an acoustic-like
propagating mode with a sound velocity exceeding the hydrodynamic
value, a signature of a relaxation process active over the probed
timescale. A quadratic dependence of the sound attenuation on the
mode frequency is observed all over the explored region,
generalizing the $Q^{2}$ law observed at small $Q$ in other
glasses.

The experiment was performed at the high resolution inelastic
scattering beamline (BL35XU) \cite{bar_spring8} of SPring-8 in
Hyogo prefecture, Japan.  High resolution was obtained using the
(11 11 11) reflection of perfect silicon crystals while a
backscattering geometry ($\frac{\pi}{2} - \theta_B \approx 0.3$
mrad) was used (for both monochromator and analyzers) in order to
obtain large angular acceptance.  A grazing incidence geometry for
the backscattering monochromator was crucial to avoid heat-load
broadening of the resolution due to the x-ray power load.  The
flux onto the sample was $\approx 3 \times 10^{9}$ photons/sec
(100 mA electron beam current) in a 0.8 meV bandwidth at 21.747
keV and a spot size of 80 $\mu m$ diameter FWHM (full width at
half maximum). The use of 4 analyzers crystals, placed with 0.78
degree spacing on the $10$ m two-theta arm (horizontal scattering
plane), and 4 independent detectors, allowed collection of 4
momentum transfers simultaneously. Slits in front of the analyzer
crystals limited their acceptance to 0.24 nm$^{-1}$ in the
scattering plane. The over-all resolution of the spectrometer was
then 1.5 to 1.8 meV, depending on the analyzer crystal. Typical
data collection times were 200 s/bin, where the bin size was fixed
at $0.25$ meV.

Elemental Selenium (purity 99.999 \%) was purchased from Cerac
Inc. The 3 mm Se shots were placed on a temperature controlled
flat surface which was heated to few degrees above the melting
point of the crystal ($T_m = 494$ K). Final thick-film samples
were obtained by applying to the viscous melt a mechanical press
and fast cooling far below the glass transition temperature. The
thickness of the Se specimens controlled by the pressure strength,
lies between 50-100 $\mu$m, which is very close to the optimum for
matching the absorption length of the X-rays ($\mu ^{-1}=50$
$\mu$m). The obtained flakes where allowed to anneal near $T_g$
and the IXS experiment was performed at $T=295$ K.

\begin{figure} [h]
\centering
%\vspace{-.5cm}
%\hspace{-1.5cm}
\includegraphics[width=.5\textwidth]{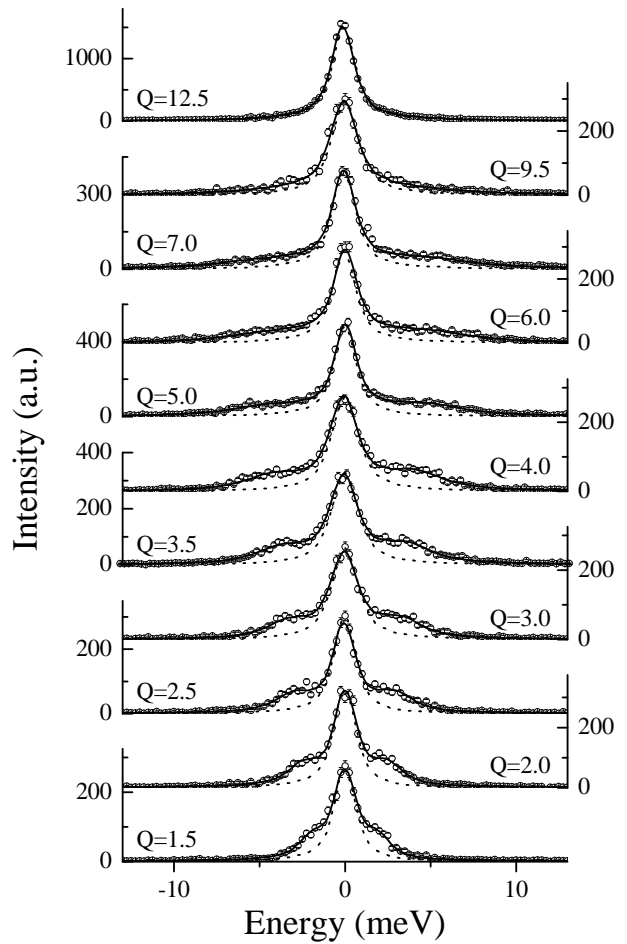}
\vspace{-.5cm}\caption{IXS spectra of glassy selenium (open dots) measured for
several momentum transfers, as indicated by the
labels (nm$^{-1}$). Also reported are the
instrument resolution (dotted line) and the best fit
lineshape according to the model discussed in the text (full
line).} \label{panel}
\end{figure}

In Fig.\ref{panel} we report the measured spectra at selected
values of (fixed) momentum transfer. Clear evidence of an
inelastic mode, at the wings of the elastic peak, can be
observed. The behavior of the energy shift of this mode vs. the
momentum transfer strongly resembles a phonon-like propagating
mode. The high inelastic/elastic ratio allows for a detailed
analysis of this mode less ambiguously than in previously studied
glasses
\cite{benassi_sio2,mas_q2,pilla_q,matic_b2o3,matic_libo,for_sio2,ruf_dens}.
Following the prescription of generalized hydrodynamics \cite{BY},
and modeling the vibrational dynamics as a Markovian process with
an instantaneous second order memory function, one ends up with a
damped harmonic oscillator line shape to represent the dynamic
structure factor $S(Q,\omega)$:

\begin{eqnarray}
\frac{S(Q,\omega)}{S(Q)}=\left [ f(Q) \delta(\omega) +
\frac{1-f(Q)}{\pi} \frac{\Omega^2(Q)\Gamma(Q)}
{(\omega^2-\Omega^2(Q))^2+\omega^2\Gamma^2(Q)} \right ]
\label{DHO}
\end{eqnarray}

The parameters $\Omega(Q)$ and $\Gamma(Q)$ represent the
characteristic frequency and attenuation of the mode, $S(Q)$ is
the static structure factor, while $f(Q)$, the
non-ergodicity factor, is related to the inelastic/elastic ratio.

The measured intensity is proportional to the convolution between
the instrumental resolution $R(\omega)$ and the dynamic structure
factor modified to account for the detailed balance condition,
namely:

\begin{equation}
I(Q,\omega )=e(Q)\int d\omega ^{\prime }\frac{\hbar \omega
^{\prime }/ KT }{1-e^{-\hbar \omega ^{\prime }/ KT }}
S(Q,\omega^{\prime})R(\omega -\omega ^{\prime })  \label{convo}
\end{equation}

with $e(Q)$ being a factor accounting for the efficiency of the
analyzers, sample thickness, the atomic form factors and other
angular-dependent instrumental correction factors \footnote{$e(Q)$
does not introduce a new fitting parameters, but prevents the
determination of $S(Q)$ from the scattered intensity}. The best
fitted lineshapes, obtained with Eq.~\ref{convo} are reported in
Fig.~\ref{panel} (full lines). It is evident that our model
function, representing a single excitation model, satisfactorily
accounts for the measured response.

\begin{figure} [h]
\centering
%\vspace{-.5cm}
%\hspace{-1.5cm}
\includegraphics[width=.5\textwidth]{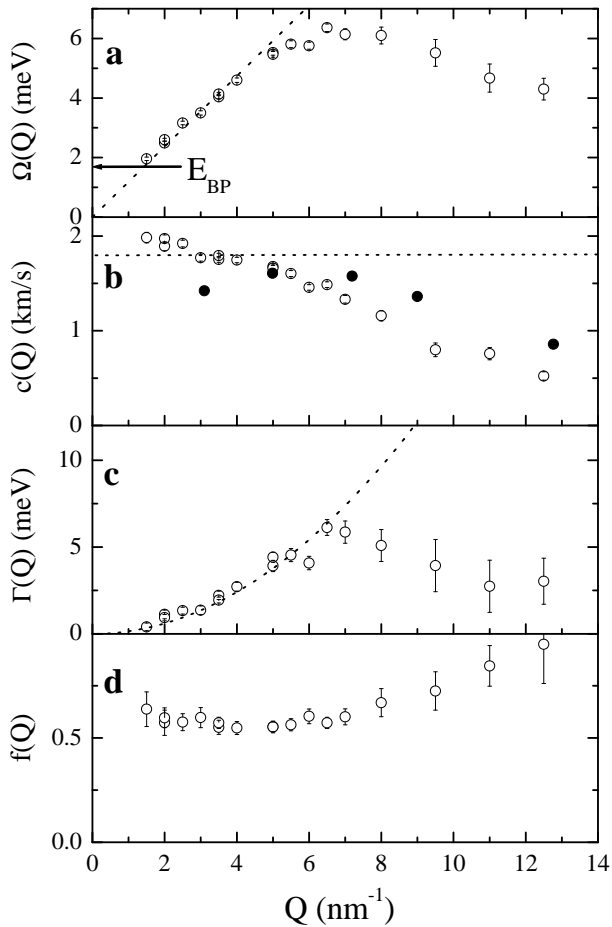}
\vspace{-.5cm}\caption{Relevant parameters obtained by fitting the
IXS measurements. (a) Maximum of the longitudinal current
correlation function (open dots) and the hydrodynamic dispersion)
\cite{and_se} (dotted line). (b) Apparent sound velocity,
$\Omega(Q) / Q $ (open dots), molecular dynamics results
\cite{gar_se} (full dots), and the hydrodynamic value (dotted
line). (c) Sound attenuation (open dots) with the best $Q^2$ fit:
$\Gamma= 0.15Q^2$ with $\Gamma$ and $Q$ expressed in meV and
nm$^{-1}$, respectively (dotted line). (d) Non-ergodicity
parameter as obtained from the inelastic/elastic ratio}
\label{results}
\end{figure}

\begin{figure} [h]
\centering
%\vspace{-.5cm}
%\hspace{-1.5cm}
\includegraphics[width=.5\textwidth]{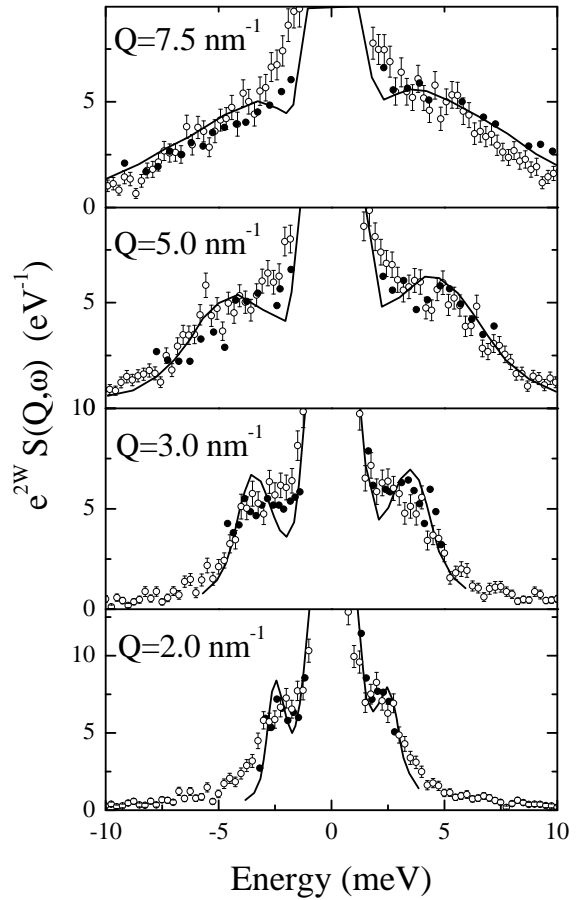}
\vspace{-.5cm}\caption{Comparison between the IXS results (open
dots and error bar) and previous INS measurements from Ref.
\cite{for_se}. Also reported is the best fit to the INS data
(full line)} \label{foret_98}
\end{figure}

The results of the present IXS experiment are summarized in
Fig.~\ref{results}, where we report the $Q$-dependence of the
three shape parameters together with that of the apparent sound
velocity $c(Q)$=$\Omega(Q)/Q$. Figure \ref{results}a) presents
compelling evidence for the existence of an acoustic-like
longitudinal branch, which is well defined up to the higher
investigated momentum transfer ($Q=12.5$nm$^{-1}$). No evidence of
localization is found at $Q$=3 nm$^{-1}$ as it has been suggested
in Ref.\cite{for_se}. Moreover, the dispersion curve extends well
beyond the boson peak frequency ($E_{BP}\approx 1.7$ meV in g-Se
at $T=300$ K \cite{and_se}), confirming earlier observations
reported in other glasses \cite{set_sci}.

The data from the INS experiment \cite{for_se} have been analyzed
assuming the simultaneous presence, in the same $Q$ region under
investigation here, of a longitudinal acoustic (LA) mode and a
local optic-like mode. The model that was utilized to describe the
LA mode implied the presence of a crossover between a propagating
and localized regime, and this crossover was found to be located
around $Q \approx 3$ nm$^{-1}$ \cite{for_se}. In contrast, MD
simulations have shown the existence of a well defined dispersion
curve up to the edge of the first pseudo-Brillouin zone ($Q\approx
10$nm$^{-1}$). This is shown in Fig.\ref{results}b, where the
sound velocity, $c(Q)$, derived by MD (full dots) and by the
present work (open dots) are reported. In Fig. \ref{foret_98}, we
plot the data from the INS experiment \cite{for_se} (full dots)
alongside with our IXS data (open dots) with the latter scaled to
optimize the overlap in the common frequency range. As can be
observed, the two sets of data are in reasonable agreement. The
small differences at the base of the elastic peak are expected
given the difference in INS and IXS resolution functions: while
both have $\approx$ 1.5 meV FWHM, the IXS resolution is
approximately Lorentzian, while the INS resolution is Gaussian.
The INS measurements, even pushing the limit of the technique at
its maximum, span only a limited range in energy due to kinematic
restrictions. This leads to best-fit lineshapes which exaggerate
the 'humps' between zero to five meV (full line in Fig.
\ref{foret_98}) and ultimately do not reproduce the IXS spectra,
measured over a much wider energy region ($-25<E<25$ meV).

The extrapolated limit of the apparent sound velocity
$\lim_{Q\rightarrow 0} c(Q)\approx 2000$ m/s
(Fig.~\ref{results}b), as measured by IXS, falls above the
hydrodynamic value ($c_0=1800$ m/s), suggesting the presence of a
possible mild positive dispersion effect. As recently reported
for another system \cite{sio2expposdisp}, this effect may be due
to a residual relaxation process -active in the glass- and with
relaxation time in the ps range. The origin of this process is
not clear at present, it could tentatively be associated with the
presence of topological disorder which modifies the hydrodynamic
sound properties at wavelength comparable with the mean
interparticle distances \cite{gcr_prlsim}.

\begin{figure} [h]
\centering
%\hspace{-1.5cm}
\includegraphics[width=.5\textwidth]{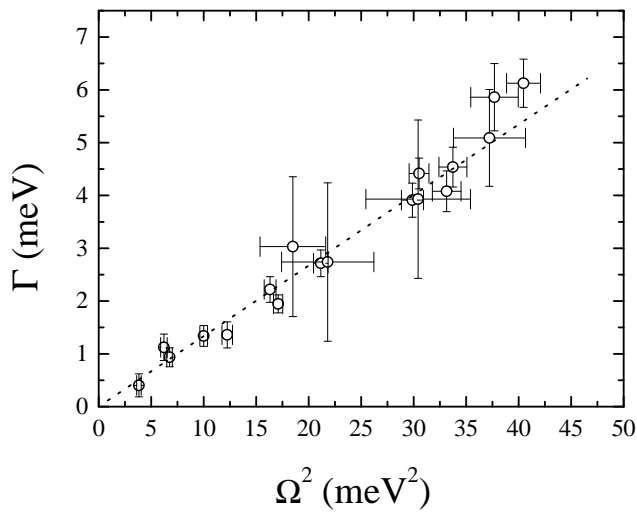}
\vspace{-6cm}\caption{Sound damping vs. square of the excitation
frequency. The quadratic dependence on the wavevector $Q^2$ shown
in Fig. \ref{results}c turns out to be the low $Q$ limit of the
more general dependence shown here. \label{gw2}}
\end{figure}

The sound attenuation parameter $\Gamma(Q)$ (Fig.~\ref{results}c)
is found to be compatible with a $Q{^2}$ law, up to $Q$'s value
where the dispersion relation is linear (i.~e. $c(Q)\approx$
constant) ($Q<5$nm$^{-1}$). This coincidence suggests a {\it
common} breakdown of the linear and quadratic laws for $\Omega(Q)$
and $\Gamma(Q)$ respectively which is expected to be caused by
structural effects, i.e. by the $Q-$ dependence of the static
structure factor \cite{BY}. In order to remove this structural
dependence we report in Fig.~\ref{gw2} the value of sound
attenuation parameter plotted against the excitation frequency for
all measured $Q$ values. As can be observed a simple power law
relation now extends up to the whole explored momentum region. The
best fitted law turns out to be $\Gamma \propto \Omega^{\alpha}$
with $\alpha$=2.15$\pm$0.10. The present observation spotlights
another aspect which needs to be encompassed in the explanation of
the ubiquitous
\cite{benassi_sio2,mas_q2,pilla_q,matic_b2o3,matic_libo} $Q^2$ (or
$\Omega(Q)^2$) dependence which is not yet fully explained.

Finally we emphasize how the combination of IXS technique and
generalized hydrodynamics provides a powerful tool to extract
information about the non-ergodicity factor $f(Q)$, i.~e. the long
time plateau of the intermediate scattering function once the
structural arrest is attained. This quantity is reported
Fig.~\ref{results}d. It should be stressed that $f(Q)$ is a true
"shape" parameter (i.~e. it does not depend on the absolute
intensity, a quantity more difficult to access experimentally)
and, in particular, it is related to the elastic-to-inelastic
intensity ratio. g-Se is found to have a remarkably low $f(Q)$
value when compared to other glasses at $T_g$, a feature which
allow a reliable lineshape analysis due to the extremely favorable
inelastic/elastic signal. This low $f(Q)$ value of selenium
mirrors the high fragility, according to the Angell classification
scheme \cite{ang2}, in agreement with recent findings that link
the value of the fragility with that of $f(Q \rightarrow 0)$ at
$T_g$ \cite{frag_scop}.

In conclusion, we presented an accurate measurement of the dynamic
structure factor in g-Se, taking advantage of the state of the art
capabilities of a new IXS facility. The inelastic/elastic signal
is particularly favorable in this system. Therefore, exploiting
the lack of kinematic constrains and the very good statistics, we
have been able to perform a robust lineshape analysis in an
extended and previously unexplored momentum-energy region. We
found evidence for a well defined longitudinal acoustic mode
extending all the way beyond the first pseudo-Brillouin zone,
characterized by dispersion and attenuation properties typical of
other glass forming materials. In particular, the boson peak
frequency lies in the linear dispersion region, thus suggesting a
link between this universal feature of glasses and the reported
high frequency acoustic excitation. Moreover, the presence of such
a well defined mode allows to get new insight into the high
frequency sound attenuation issue. Specifically, a quadratic
dependence of the sound attenuation on the mode frequency extends
the $Q^{2}$ law holding in a smaller range and observed in other
glass formers. Such an observation indicates the prominent role of
the excitation frequency rather than of its wavevector, and
certainly calls for further investigations.

The synchrotron radiation experiment was performed at the SPring-8
with the approval of the Japan Synchrotron Radiation Research
Institute (JASRI) (Proposal No. 2003A0357-ND3-279).

%\bibliographystyle{apsrev}
%\bibliography{../glass}

\end{document}